\documentclass[preprint,showpacs,preprintnumbers,amsmath,amssymb]{revtex4-1}

\usepackage{graphicx}% Include figure files
\usepackage{dcolumn}% Align table columns on decimal point
\usepackage{bm}% bold math
\usepackage{color}
\usepackage{upgreek}

\begin{document}

\title{Carrier Confinement in GaN/Al$_x$Ga$_{1-x}$N Nanowire Heterostructures for $0 < x \leq$ 1}% Force line breaks with \\

\author{Florian Furtmayr$^{a,b}$}
\author{J\"org Teubert$^a$}\email{joerg.teubert@exp1.physik.uni-giessen.de} 
\author{Pascal Becker$^a$}
\author{Sonia Conesa-Boj$^c$}
\author{Joan Ramon Morante$^{c,d}$}
\author{Jordi Arbiol$^e$}
\author{Alexey Chernikov$^f$}
\author{S\"oren Sch\"afer$^f$}
\author{Sangam Chatterjee$^f$}
\author{Martin Eickhoff$^a$}

\affiliation{$^a$ I. Physikalisches Institut, Justus-Liebig-Universit\"at Gie\ss en, Heinrich-Buff-Ring 16, D-35392 Gie\ss en, Germany}
\affiliation{$^b$ Walter Schottky Institut, Technische Universit\"at M\"unchen, Am Coulombwall 4, D-85748 Garching, Germany}
\affiliation{$^c$ Dept. d'Electrònica, Universitat de Barcelona, c/ Marti Franques 1, 08028 Barcelona, CAT, Spain}
\affiliation{$^d$ IREC, Catalonia Institute for Energy Research, 08019 Barcelona, CAT, Spain}
\affiliation{$^e$ ICREA and Institut de Ciencia de Materials de Barcelona, CSIC, Campus de la UAB, 08193 Bellaterra, CAT, Spain}
\affiliation{$^f$ Faculty of Physics and Materials Science Center, Philipps Universit\"at Marburg, Renthof 5, D-35032 Marburg, Germany}

\date{\today}% It is always \today, today,
             % but any date may be explicitly specified

\begin{abstract}
The three dimensional carrier confinement in GaN nanodiscs embedded in GaN/Al$_x$Ga$_{1-x}$N nanowires and its effect on their photoluminescence properties is analyzed for Al concentrations between x = 0.08 and 1. Structural analysis by high resolution transmission electron microscopy reveals the presence of a lateral AlGaN shell due to a composition dependent lateral growth rate of the barrier material. The structural properties are used as input parameters for three dimensional numerical simulations of the confinement which show that the presence of the AlGaN shell has to be considered to explain the observed dependence of the emission energy on the Al concentration in the barrier. The simulations reveal that the maximum in the emission energy for $x \approx 30\%$ is assigned to the smallest lateral strain gradient and consequently the lowest radial internal electric fields in the nanodiscs. Higher Al-concentrations in the barrier cause high radial electric fields that can overcome the exciton binding energy and result in substantially reduced emission intensities. Effects of polarization-induced axial internal electric fields on the photoluminescence characteristics have been investigated using nanowire samples with nanodisc heights ranging between 1.2\,nm and 3.5\,nm at different Al concentrations. The influence of the quantum confined Stark effect is significantly reduced compared to GaN/AlGaN quantum well structures which is attributed to the formation of misfit dislocations at the heterointerfaces which weakens the internal electric polarization fields. 
\end{abstract}


% 81.07.Gf: Materials science / nanoscale materials / nanowires
% 61.46.-w: Structure of nanoscale materials
% 81.07.-b: Nanoscale materials and structures: fabrication and characterization 
% 78.67.Uh: Optical properties of low-dimensional, mesoscopic, and nanoscale materials and structures / nanowires
% 78.20.Bh: Optical properties, condensed-matter spectroscopy and other interactions of radiation and particles with condensed matter / Theory, models, and numerical simulation

\maketitle

\section{Introduction}
GaN nanowires (NWs) can be grown by molecular beam epitaxy using a catalyst free growth process under nitrogen-rich growth conditions \cite{Calleja2000, Calleja2007}. Their low defect density makes them an ideal model system for the investigation of basic material properties, such as the effect of incorporation of dopants on the structural and optical properties \cite{Furtmayr2008, Furtmayr2008a, Arbiol2009} or limited size effects such as the influence of surface band bending on the conductivity \cite{Calleja2007a, Calleja2007} or the decay time of persistent photocurrent \cite{Calarco2005}.

In addition, due to their low density of structural defects group III-nitride NWs and nanowire heterostructures (NWH) are considered as a promising approach for the realization of improved nano- or optoelectronic devices \cite{Calle1997, SanchezGarcia1998, Calleja1999}. With this respect, the realization of heterostructures embedded in nanowires is of major importance and has been demonstrated by different groups \cite{Yoshizawa1997, Ristic2005, Rivera2007, Calarco2005, Kikuchi2004}. The optical properties of such heterostructures are determined by carrier confinement due to reduced lateral and axial dimensions and by the presence of polarization-induced internal electric fields. For the latter both spontaneous and piezoelectric polarization effects have to be taken into account, as in the case of heterostructures pseudomorphically grown in NWs neither the GaN quantum wells (here referred to as nanodiscs (NDs)) nor the AlGaN barriers can be assumed to be free of strain.

The luminescence properties of GaN NDs embedded in GaN/AlGaN or GaN/AlN NWs have been investigated in different works and strain-induced piezoelectric polarization has been found to be of major importance for the interpretation of the observed results. In ref.~\cite{Ristic2005a}, non-uniform strain along the ND diameter has been considered to be responsible for broadening of the ND photoluminescence (PL) as it affects both the local band gap and the internal polarization fields. It was concluded that this results in variations of the transition energy up to 100\,meV. Model calculations based on a one-dimensional Schr\"odinger-Poisson solver have revealed that the radial strain profile determines the carrier confinement and can cause luminescence quenching that was observed for very thin NDs. In ref.~\cite{Rivera2007} these calculations have been improved and the existence of strain-confined states with spatial separation of carriers has been proposed. Carrier localization in the ND center was found to be improved with increasing ND height.
Renard et al.~have observed a decrease of the GaN/AlN ND emission energy in NDs below the band gap of GaN with increasing height ranging from 1.5\,nm to 4.5\,nm in radial direction \cite{Renard2009}, less pronounced than what has been reported for GaN/AlN thin film quantum wells \cite{Adelmann2003}. This behavior was partially attributed to elastic relaxation of the GaN NDs at the sidewalls. Partial strain relaxation has been confirmed in ref.~\cite{Landre2010} for the case of GaN NDs with a height of 2\,nm between AlN barriers of 2.3\,nm by in-situ x-ray diffraction and HRTEM. For that case it was found that the in plane lattice parameter of the GaN NDs is 3.15\,{\AA} corresponding to an alloy ratio of Al$_{0.55}$Ga$_{0.45}$N, determined by the almost identical thickness of NDs and barriers.

The strain profile along the ND diameter and the resulting carrier confinement strongly depend on the boundary conditions, i.e.~the strain state of the ND at the NW sidewalls, which has been assumed to be fully relaxed in the above mentioned reports. However, it should be mentioned that the presence of a thin AlN shell on the NW sidewalls due to lateral growth rate of AlN was observed by the same groups \cite{Ristic2005, Renard2009}. In a recent report of Zagonel et al.~a strong dependence of the  dispersion of the emission energy of AlN/GaN NDs with different heights along one single nanowire was observed by spatially resolved cathodoluminescence and the variation of the emission energy was attributed to the presence of a lateral AlN shell \cite{Zagonel2011}. 

In the present paper we analyze the carrier confinement in GaN NDs embedded in GaN/AlGaN nanowires by investigating the impact of the Al-concentration in the $\text{AlGaN}$ barriers, [Al]$_\text{bar}$, and the ND height, $d_\text{ND}$, on the optical transition energies. The results of the optical analysis by photoluminescence spectroscopy (PL) are compared to numerical calculations based on input parameters that are obtained from structural analysis by high resolution transmission electron microscopy (HRTEM). In these calculations, the lateral growth rate of the barrier material on the NW sidewalls is taken into account. The transition energies are calculated as a function of $\text{[Al]}_\text{bar}$ and compared to experimental data, allowing to identify the influence of different structural parameters on the ND optical properties. The presence of misfit dislocation-type defects is obtained from detailed structural analysis by HRTEM and their effect on the polarization-induced internal electrical field and ND strain relaxation is discussed. 
Time resolved photoluminescence spectroscopy (TR-PL) indicates the presence of internal strain-induced radial electric fields that cause dissociation of excitons for $\text{[Al]}_\text{bar} > 0.3$.

\section{Experimental and Sample Details}
\label{ExpSample}
GaN NDs in NWs were grown by plasma assisted molecular beam epitaxy (PAMBE) on Si(111) substrates. Ga and Al were supplied by thermal effusion cells, whereas nitrogen was introduced by a RF-plasma source. Nitrogen-rich growth conditions ($\text{V/III} \approx 4$ compared to a GaN layer) were applied to form self assembled NWs. Details on the growth can be found in ref.~\cite{Furtmayr2008}. First, a GaN NW base with a length of 350\,nm to 400\,nm and a diameter between 25\,nm and 50\,nm was grown, followed by a nine-fold GaN ND structure ($d_\text{ND}$ was varied from 1.2\,nm to 3.5\,nm) between barriers of 7\,nm AlN or Al$_x$Ga$_{1-x}$N and a 20\,nm cap layer consisting of the barrier material (all dimensions determined by HRTEM analysis). The substrate temperature was kept at 775\,$^\circ$C for all parts of the NW structure to suppress the formation of a two dimensional  GaN layer connecting individual NWs which has been reported e.g.~in ref.~\cite{Stoica2008}. 
%For the growth temperatures applied here, two dimensional film growth between the NWs was not observed \cite{Furtmayr2008}. 
The temperature of the Ga effusion cell was kept at 1012\,$^\circ$C, corresponding to a beam equivalent pressure of BEP$_\text{Ga} = 3.9 \times 10^{-7}\,\text{mbar}$, during growth of both GaN base and Al$_x$Ga$_{1-x}$N barriers. For the latter Al was additionally supplied at cell temperatures between 1020\,$^\circ$C and 1170\,$^\circ$C. For the formation of AlN barriers, the Al cell was operated at a temperature of 1185\,$^\circ$C (BEP$_\text{Al} = 3.5 \times 10^{-7}\,\text{mbar}$). The different BEP values and corresponding Al-concentrations are summarized in Table~\ref{tab:GrowthDetails}. The Al content was determined by two independent methods: Firstly, directly from the BEP$_\text{Al}$ to BEP$_\text{Ga}$ ratio measured by a Bayard-Alpert flux gauge. Secondly, by evaluating the position of the PL emission peak of specific reference GaN/AlGaN nanowire heterostructures without NDs in which the GaN base region was directly followed by growth of 100\,nm AlGaN, assuming a bowing parameter of $b = 1.3\,\text{eV}$ (ref.~\cite{Angerer1997}). In the following, the [Al]$_\text{bar}$-values according to the BEP ratio in Table \ref{tab:GrowthDetails} are used, which are in good agreement with the optical measurements.

% Table 1

To realize multi ND samples with a reduced AlGaN shell thickness for comparison, samples were grown with a reduced Ga-flux during growth of the ND part, resulting in a reduction of the total metal flux and a strong reduction of the lateral growth. 

Structural characterization by high resolution transmission electron microscopy (HRTEM), scanning transmission electron microscopy (STEM), high annular angular dark field microscopy (HAADF), and electron energy loss spectroscopy (EELS) was carried out to extract radial and axial growth rates and to obtain structural input parameters for the numerical simulations. In addition, it was used to analyze deviations from the ideal nanostructure in terms of dislocation formation, relaxation, etc.. Experiments were carried out in a JEOL2010F microscope equipped with a GATAN GIF EELS spectrometer. For microscopy analysis the NWs were mechanically removed from the substrate and transferred onto a holey carbon TEM grid as described in ref.~\cite{FontcubertaiMorral2007}.

The optical properties of NW ensembles were studied by low temperature ($T=4\,\text{K}$) photoluminescence spectroscopy. The as-grown samples were mounted in an Oxford cryostat and the PL emission was collected in backscattering geometry. To allow comparison, all spectra shown in this work were recorded using a frequency-quadrupled pulsed Nd:YAG laser (CryLas) at 266\,nm for excitation and a Spex 0.22\,m spectrometer for light dispersion. 

For comparison, a PL-setup equipped with a HeCd cw-laser (325\,nm, 30\,mW) and a 1\,m Jobin Yvon monochromator with a photomultiplier tube (Hamamatsu R375, $160 - 850\,\text{nm}$) was used for samples with an emission energy below 3.8\,eV. Spectra that were recorded using the pulsed excitation show emission energies that are systematically higher by up to 110\,meV due to partial screening of internal electric fields. However, all trends discussed in this work could be reproduced for both excitation sources. 

Time-resolved photoluminescence measurements were performed using a pulsed 100\,fs - Ti:sapphire laser at 80\,MHz repetition rate as an excitation source. The pump power was set to 1\,$\upmu$W, corresponding to a photon flux at the sample surface of about $10^9\,\text{cm}^{-2}$ per pulse. The PL signal was detected in a streak camera setup with spectral and temporal resolution of 1\,nm and 20\,ps, respectively.

\section{Experimental Results}

\subsection {Structural Analysis}
\label{TEM}

\subsubsection*{Geometric structure of GaN nanodiscs in GaN/AlGaN nanowires}

\begin{figure}[tb]
	\includegraphics[trim=0.0cm 0.0cm 0.0cm 0.0cm, width=8cm]{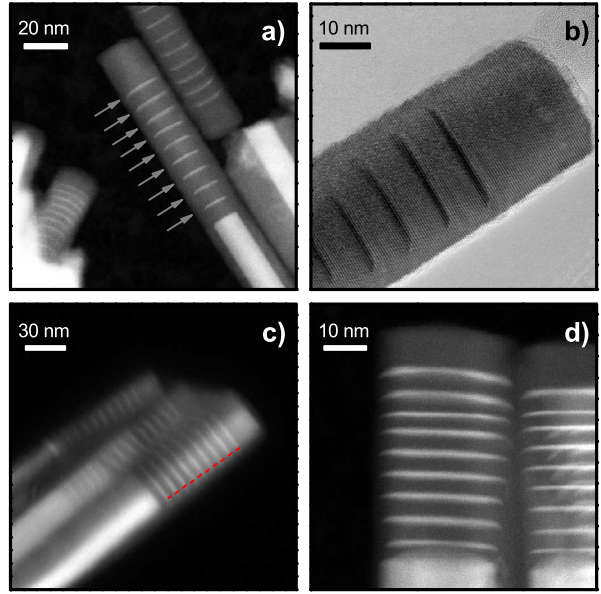}
    \caption{
(Color online) 
a) HAADF (high angle annular dark field) STEM image of nanowires with 1.7\,nm thick GaN nanodiscs surrounded by AlN barriers. The GaN appears in bright contrast, the surrounding AlN in dark.
b) 	HRTEM image (bright field) of the same sample. The increase of ND diameter along the growth direction is visible. 
c) 	HAADF STEM of a NW with $\text{[Al]}_\text{bar} = 0.41$. Lateral growth is observed here as well, however with a reduced rate compared to AlN-barriers. The dotted line marks the increase in GaN diameter along the growth axis.
d) 	HAADF image of a sample with $\text{[Al]}_\text{bar} = 0.72$ grown under modified growth conditions as described in section \ref{ExpSample}. The lateral growth is efficiently reduced in this case. As an upper estimate we obtain 2\% of the axial growth rate in that case which corresponds to a reduction by approximately a factor of four compared to the growth conditions applied in a)-c).}
\label{fig:Fig01}
\end{figure}

The geometric structure of the NWH samples was determined by HRTEM analysis. In Figure~\ref{fig:Fig01}a a HAADF STEM image of a sample with nine NDs of $d_\text{ND} = 1.7\,\text{nm}$ and 7\,nm AlN barriers is shown (cf.~bright field HRTEM image of the same sample in Figure~\ref{fig:Fig01}b). In the GaN/Al$_x$Ga$_{1-x}$N structures with high [Al]$_\text{bar}$ the presence of a lateral AlN or AlGaN shell (also reported in refs.~\cite{Ristic2005, Tchernycheva2008, Zagonel2011} for AlN barriers) was observed (cf.~Figure~\ref{fig:Fig01}a,~b,~c). Quantitative analysis by HRTEM reveals a radial growth rate of 11\% of the axial growth rate for AlN and of $(5\pm 2)$\% for the case of Al$_{0.41}$Ga$_{0.59}$N barriers. In ref.~\cite{Tchernycheva2008} a higher lateral growth rate of 35\% of the axial growth rate was determined for AlN. We attribute this difference to a lower substrate temperature during deposition used in that work. The lateral growth rate was found to be related to the total metal flux. For constant Ga-flux it is directly related to the Al-content and, based on the values mentioned above, we assume a linear dependence of the lateral growth rate on [Al]$_\text{bar}$, depending on the deposition temperature and on the exact III-V ratio during deposition. In contrast, if the total metal flux is reduced during barrier deposition (i.e.~modified growth conditions described in section \ref{ExpSample}), the lateral growth is strongly reduced as shown in Figure~\ref{fig:Fig01}d for a GaN/Al$_{0.72}$Ga$_{0.28}$N NWH with $d_\text{ND} = 2.5\,\text{nm}$. As an upper estimate for the lateral growth rate we obtain 2\% of the axial growth rate for $\text{[Al]}_\text{bar} = 0.7$ which corresponds to a reduction by approximately a factor of four compared to the previously mentioned growth conditions. 

% Figure 1   **************************************************************************************************

The present HAADF STEM results do not allow a precise determination of the shell composition. From the viewpoint of adatom kinetics, well studied on polar GaN surfaces \cite{Zywietz1998, Adelmann2002, Iliopoulos2002} one could expect that the Al-content in the shell is significantly higher than in the barriers. However, the experiments did not reveal a significant difference in contrast of the shell and the respective barrier material. Considering the quadratic dependence on the Z-number we have to assume that the shell consists of the barrier material.  

It is important to notice that the Al-induced lateral growth has two major implications on the properties of the NW heterostructures: (i) A gradual decrease of the lateral shell thickness within one NW due to lateral expansion of the NW diameter during growth of each barrier (cf.~Figure~\ref{fig:Fig01}a,~b,~c). (ii) An effective three dimensional confinement of carriers in the GaN NDs, gaining in importance with increasing $\text{[Al]}_\text{bar}$, i.e.~with increasing lateral AlGaN growth rate.

As observable in Figure~\ref{fig:Fig01}b and \ref{fig:Fig01}d, the GaN NDs appear as flat discs with sharp interfaces to the barriers, though monolayer fluctuations of the ND height are observed \cite{Rigutti2010a}. On the outer edges the discs are slightly bent downwards, as also mentioned in ref.~\cite{Ristic2005, Bougerol2010}. Comparison of NDs in different NWs reveals that this structure originates in faceting of the GaN base top surface, where $\{ 1 \bar{1} 0 3\}$ planes form the outer edges (c.f.~Figure~\ref{fig:Fig01}b,~d). Subsequent overgrowth with ND and barrier material, including lateral growth of the latter, results in the typical shape presented in Figure~\ref{fig:Fig01}a,~b for a GaN/AlN NWH and in Figure~\ref{fig:Fig01}c for a GaN/Al$_{0.41}$Ga$_{0.59}$N NWH. If the total metal flux during barrier growth is reduced according to the procedure described in section \ref{ExpSample}, the reduced lateral growth rate ensures an enhanced conformal reproduction of the GaN base facets, appearing as more pronounced downward bending of the ND edges, shown in Figure~\ref{fig:Fig01}d. 

\begin{figure}[tb]
	\includegraphics[trim=0.0cm 0.0cm 0.0cm 0.0cm, width=8cm]{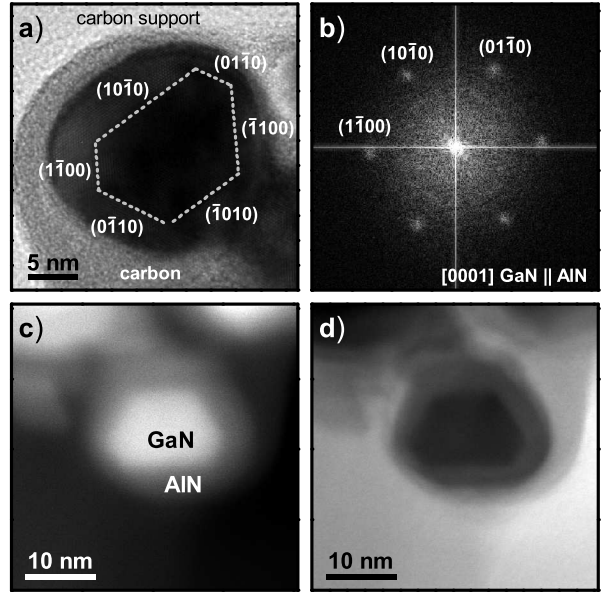}
    \caption{
TEM cross-section of a GaN NW surrounded by the AlN shell. The GaN core forms a hexagonal prism section with $\{ 1 0 \bar{1} 0 \}$ planes (m-planes) as lateral facets. The surrounding AlN shell is more cylindrical with rounded facets. 
a)	HRTEM image with GaN core sidewalls (dotted lines) indexed. 
b)	Corresponding selected area electron diffraction (SAED) pattern, proving that the sidewalls are m planes.
c)	HAADF image.
d)	Bright field image.
}
\label{fig:Fig02}
\end{figure}

In Figure~\ref{fig:Fig02} TEM cross-sections of an GaN/AlN ND sample are shown, demonstrating the presence of the lateral AlN shell. The GaN core forms a hexagonal prism with $\{ 1 0 \bar{1} 0\}$ planes (m-planes) as lateral facets. The difference in the cross section of the prismatic GaN core of the NW and the rounded shell directly reflects the difference in surface kinetics for Ga and Al. Lymperakis et al.~have recently reported that the anisotropic growth rate of c-axis oriented GaN NWs is due to anisotropic surface thermodynamics and nucleation \cite{Lymperakis2009}. From this point of view the present results demonstrate that the anisotropy for the incorporation probability of Al-adatoms during AlN-growth under the growth conditions applied here is less pronounced.

% Figure 2 ********************************************************************************************************

Contrast changes in bright field STEM images (not shown here) indicate the presence of axial strain inhomogeneities also inside the GaN base. Corresponding electron energy loss spectroscopy (EELS) line scans measured along the NW axis reveal an increase of the Al signal from the NW base to the top with a typically low amount of Al being in the lower 100\,nm, representing an increasing shell thickness. This observation can be explained using a geometrical argument recently suggested by Foxon et al.~\cite{Foxon2009}. Following those considerations and assuming a NWs density of $250\,\upmu\text{m}^{-2}$ (i.e.~an average distance of 50\,nm between neighboring NWs \cite{Furtmayr2008}) and an average height of 350\,nm when the first AlGaN barrier is grown, one expects a lateral growth rate of 18\% assuming a sticking coefficient of 1 for Al adatoms (in the present MBE system the impinging angle of adatoms is $30^\circ$). As shielding by neighboring NWs can further decrease the lateral growth rate the obtained results show reasonable agreement to the model in ref.~\cite{Foxon2009}.

\subsubsection*{Dislocations}

\begin{figure}[tb]
	\includegraphics[trim=0.0cm 0.0cm 0.0cm 0.0cm, width=8cm]{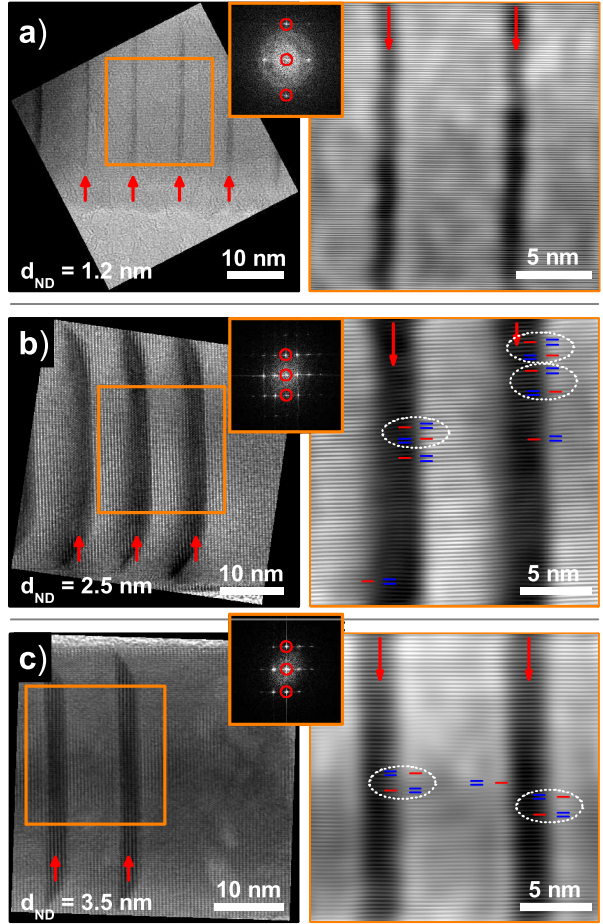}
    \caption{
(Color online) HRTEM, frequency filtering on the $[1\bar{1}00]$ lateral planes parallel to the growth axis and dislocation analysis of GaN/AlN NWH samples with a ND height of $d_\text{ND} = 1.2$, 2.5 and 3.5\,nm, see a), b) and c), respectively. NDs have been marked with red arrows. The presence of misfit dislocations has been marked with blue and red lines. Most dislocations appear pair wise thus they are successfully compensated by an inverse dislocation as marked by white dotted ellipsoids. Single dislocations are observed with a lower density.
}
\label{fig:Fig03}
\end{figure}

To clarify the presence of misfit dislocations structural analyses by HRTEM were carried out. GaN/AlN NWHs were analyzed by means of HAADF STEM to allow a clear identification of ND and barrier regions composed of materials with different Z (Figure~\ref{fig:Fig01}). For comparison we have investigated three GaN/AlN ND structures with different ND heights containing a lateral shell with an axial gradient according to a radial growth rate of 11\% from the axial growth rate. We have analyzed the local atomic structure by filtering the power spectra obtained on the HRTEM micrographs to enhance the contrast on the $[1\bar{1}00]$ lateral planes parallel to the growth axis and to analyze the possible presence of misfit dislocations (Figure~\ref{fig:Fig03}).

%Figure 3 ******************************************************************************************+

In the sample with a ND height of $d_\text{ND} = 1.2\,\text{nm}$ no indications for the presence of misfit dislocations were found (Figure~\ref{fig:Fig03}a), meaning that the GaN NDs are compressively strained along the $[1\bar{1}00]$ planes, as expected. However, for an increased ND height of $d_\text{ND} = 2.5\,\text{nm}$ misfit dislocations were observed in the ND region, as shown in Figure~\ref{fig:Fig03}b. For $d_\text{ND} = 3.5\,\text{nm}$ this trend was confirmed (Figure~\ref{fig:Fig03}c). It should be mentioned that most of these dislocations occur pairwise, being compensated by an inverse dislocation in the close vicinity as marked by dashed circles in Figure~\ref{fig:Fig03}b,~c. The presence of these pair dislocations has also been observed in GaN/AlN quantum well structures \cite{Kandaswamy2009}. Single dislocations, not being compensated by an inverse one, are also observed. As can be extracted from Figure~\ref{fig:Fig03}, the absolute number of such dislocations in one ND is low. However, assuming the presence of only two single dislocations along a ND diameter of 40\,nm implies an areal density of $5 \times 10^{10}\,\text{cm}^{-2}$ and allows almost complete relaxation of the in-plane lattice mismatch.

The presence of such defects is in apparent contradiction to the results presented in ref.~\cite{Landre2010}. In that report multi ND structures consisting of 2\,nm GaN NDs in 2.3\,nm AlN barriers on a 200\,nm GaN NW base were investigated and no formation of dislocations was observed. Compared to that work the thickness of the AlN barriers in the present experiments is increased by a factor of three, resulting in an increased compressive strain inside the NDs. In addition, the formation of a lateral AlN shell in the samples investigated here, which suppresses relaxation at the lateral surfaces, leads to a reduction of the critical layer thickness compared to calculations in ref.~\cite{Glas2006} and was not observed in the NWHs investigated in ref.~\cite{Landre2010}. In contrast, Bougerol et al. found dislocations in GaN/AlN NWHs at the oblique sides (at the outer edges) of 2.5\,nm thick GaN NDs separated by 12\,nm AlN barriers (i.e.~for a similar ratio of ND to barrier thickness as used in the present experiments) \cite{Bougerol2010}. In that case the NWHs exhibited a pronounced lateral AlN shell as well.

\subsection{Optical Properties}
\subsubsection*{Variation of ND height}

For NWs grown along the polar $[0001]$ direction the presence of axial piezoelectric fields and the influence of the quantum confined Stark effect (QCSE) is expected. In Figure~\ref{fig:Fig04}a, the low temperature ($T = 4\,\text{K}$) emission of GaN/Al$_x$Ga$_{1-x}$N NWH samples with $\text{[Al]}_\text{bar} = 0.14$ and varied $d_\text{ND}$ is displayed. The peaks at 3.455\,eV and 3.477\,eV originate from the GaN base region and are not influenced by variations of $d_\text{ND}$. The 3.477\,eV luminescence is attributed to the recombination of donor bound excitons, whereas the 3.45\,eV emission line has been assigned to an exciton bound to a defect state at the nanowire surface \cite{Furtmayr2008a} or to defects at the interface to the Si(111) substrate \cite{Pfueller2010}. The full width at half maximum (FWHM) of the 3.477\,eV emission of $(13 \pm 2)$ meV is broadened due to the growth of the heterostructure compared to ensembles of pure GaN NWs where a FWHM of 2\,meV was found \cite{Furtmayr2008a}.

% Figure 4 *************************************************************************************************

The ND-related peak shows a red-shift from 3.587\,eV to 3.512\,eV when the ND height is increased from 1.7\,nm to 3.5\,nm, whereas the FWHM remains constant at $(41 \pm 2)\,\text{meV}$. The relative intensity of the ND emission significantly decreases for $d_\text{ND} = 3.5\,\text{nm}$. As this value exceeds the exciton Bohr radius in GaN (2.7\,nm \cite{Steude1999, Gallart2000}) we attribute this behavior to separation of the exciton by the internal axial electric field (cf.~section~\ref{discussion}). The spectra for samples with pure AlN barriers, shown in Figure~\ref{fig:Fig04}b, show a red-shift of the ND-related emission from 4.10\,eV for $d_\text{ND} = 1.2\,\text{nm}$ to 2.95\,eV, well below the band edge of GaN, for $d_\text{ND} = 3.5\,\text{nm}$. A similar behavior was recently reported by Renard et al.~for single NDs between 10\,nm thick AlN barriers \cite{Renard2009}. 

For $\text{[Al]}_\text{bar} = 1.0$ the FWHM of the ND emission peaks is determined to 410\,meV, significantly higher than for the sample with lower Al content, indicating that the dispersion of the ND emission energy in one NW is enhanced for increasing Al concentration in the barrier, as it has recently been reported for low Al concentrations in ref.~\cite{Rigutti2010a} and will be confirmed by the numerical simulations shown below. Also the contributions from the GaN base region at 3.477\,eV and around 3.27\,eV, that do not shift with the well thickness, are broadened.

In spite of the apparent qualitative agreement with the behavior of GaN/AlGaN quantum well structures the influence of the QCSE is significantly reduced. This peculiar effect which is related to the finite size and the absence of translational symmetry in lateral dimension will be discussed in detail in section~\ref{discussion}.

\subsubsection*{Emission energy vs.~Al concentration}

In polar GaN/AlGaN quantum well structures (QWs) an increasing difference in Al concentration at heterointerfaces has two different effects on the photoluminescence properties. For low differences both the emission energy and the intensity increase due to an improvement of carrier confinement. For larger differences, the accumulation of polarization-induced interface charge results in strong internal electric fields along the growth direction that give rise to a reduction of the emission energy and intensity due to the QCSE \cite{Grandjean1999a}. 

Figure~\ref{fig:Fig05}a shows the low temperature ($T = 4\,\text{K}$) PL spectra of GaN NWH ensembles with $d_\text{ND}=1.7\,\text{nm}$ and different $\text{[Al]}_\text{bar}$. While the energy of the GaN-base emission at 3.48\,eV is independent of the barrier composition its width increases with $\text{[Al]}_\text{bar}$ which can only be justified by the influence of the AlGaN shell thickness gradient. The ND emission energy increases with $\text{[Al]}_\text{bar}$ up to 3.73\,eV for $\text{[Al]}_\text{bar} = 0.34$ and subsequently decreases down to 3.69\,eV for pure AlN barriers (cf.~full symbols in Figure~\ref{fig:Fig05}b). The FHWM of the ND emission peak, indicated by the open symbols in Figure~\ref{fig:Fig05}b, increases with $\text{[Al]}_\text{bar}$ from 28\,meV for the sample with the smallest Al-concentration to 410\,meV for the sample with AlN barriers. For low and medium $\text{[Al]}_\text{bar}$ the intensity of the NWH emission exceeds that of the GaN nanowire base by up to one order of magnitude, the maximum intensity being observed for $\text{[Al]}_\text{bar} = 0.26$. For $\text{[Al]}_\text{bar} > 0.34$ a direct comparison of the emission intensities is difficult due to the strongly increased FWHM at high $\text{[Al]}_\text{bar}$ and the fact that for $\text{[Al]}_\text{bar} > 0.55$ the excitation laser radiation at 266\,nm is not absorbed in the barriers.

% Figure 5 ************************************************************************************************

\subsubsection*{Carrier lifetimes}

In order to determine the decay time of the ND emission for selected samples at low temperature ($T = 10\,\text{K}$) time resolved photoluminescence (TR-PL) measurements were carried out. Results were obtained in a regime of low excitation density as affirmed by excitation power dependent measurements so that screening effects can be neglected. Figure \ref{fig:Fig06} presents the results for GaN/AlGaN NWH ensembles with $d_\text{ND}=1.7\,\text{nm}$ and different $\text{[Al]}_\text{bar}$ (blue filled circles, cf.~Figure \ref{fig:Fig05}). While the ND emission shows decay times between 500\,ps and 1\,ns for $\text{[Al]}_\text{bar} \leq 0.34$, the lifetime increases significantly above this threshold up to 7\,ns for $\text{[Al]}_\text{bar} = 1.0$. For later discussion of axial vs. lateral internal electric fields decay times as a function of ND height $d_\text{ND}$ for $\text{[Al]}_\text{bar} = 1.0$ are given in the insert of Figure\,\ref{fig:Fig06} (red filled squares) in comparison with corresponding results for GaN/AlGaN quantum well (QW) structures (open black squares) from ref.~\cite{Renard2009a}. The dependence on the thickness is much weaker for the ND emission compared to that of QW structures. While QWs show shorter decay times below approximately 2\,nm they exhibit much longer lifetimes above that value.

% Figure 6 ************************************************************************************************

With regard to section~\ref{discussion} it has to be noted here that the behavior of transition energy, emission intensity and carrier lifetime strongly differs from that of GaN/AlGaN QWs and cannot be described solely by an interplay of improved carrier confinement due to increased barrier heights and the influence of the QCSE due to the presence of polarization-induced electric fields. In fact, the presence of the AlGaN shell, a unique feature of NWs, and the related gradient in thickness affects the strain distribution in the NDs and impacts the PL emission energy as well as its dispersion along one NW. Additionally, it has to be emphasized that in the case of NDs in NWs the radial strain profile also causes the presence of lateral electric fields \cite{Ristic2005a, Rivera2007}. For high  Al content, the latter can cause the separation of excitons in radial direction and contribute to the decrease in luminescence intensity and the increase in carrier lifetime. To clarify the contribution of the different effects numerical simulations were carried out and will be discussed in section~\ref{discussion}.

\section{Modeling and Discussion}
\label{discussion}

The Al-induced lateral growth has major implications on the properties of the NWHs. The gradual decrease of the lateral shell within one NW due to lateral expansion of the NW diameter during growth of each barrier (Figure~\ref{fig:Fig01}b,~c) involves a modification of the strain distribution along the ND region, i.e.~each of the NDs experiences different radial and axial strain conditions. The gradient in shell thickness should result in a dispersion of the ND optical properties along one NW.  Besides these geometric effects complete embedment of the NDs in barrier material leads to an effective three dimensional confinement of photo-excited carriers in the GaN NDs, gaining in importance with increasing $\text{[Al]}_\text{bar}$. Furthermore, the effect of the related strain distribution on the three dimensional carrier confinement has to be taken into account.

Numerical simulations of the strain distribution and the carrier confinement as well as the effect on the PL properties have been reported in refs.~\cite{Ristic2005a, Rivera2007}. These works mainly focus on the strain-induced lateral variation of the band edges, discussing the possibility of lateral carrier separation as an explanation for the decreasing luminescence intensity of thin multiple NDs. The authors also discussed relaxation of in-plane compressive strain inside the GaN NDs from the center towards the sidewalls as an explanation for emission broadening in the range of 100\,meV for samples with $\text{[Al]}_\text{bar} = 0.2$ \cite{Ristic2005}. Those simulations targeted on the identification of the effect of lateral relaxation at the NW sidewalls on the carrier confinement for fixed $\text{[Al]}_\text{bar}$ and were based on zero strain boundary conditions in lateral direction and periodic axial boundary conditions (justified by the symmetric sample structure). In a more recent report by Jahn et al., lateral strain inhomogeneities in the NDs were discussed as a possible reason for line broadening of 155\,meV in GaN/Al$_{0.28}$Ga$_{0.72}$N NDs ensembles and of 80\,meV in the ND stack of a separate wire \cite{Jahn2007}. Strain variations along the NW axis have not been considered. These above mentioned reports indicate that the detailed knowledge of the strain distribution inside the NWH is of great importance for an understanding of the carrier confinement effects. 

In a theoretical work of Mojica et al.~GaN/AlN single ND structures were considered by modeling the NWs as cylindrical pillars. The authors discuss the possible partial screening of axial polarization induced electric fields due to the presence of a two dimensional electron gas  and surface states at the top NW-surface \cite{Mojica2010} leading to a reduction of the QCSE. However, the presence and influence of a surrounding lateral shell on the strain distribution and the band structure has not been taken into account. 

In general, carrier confinement in III-nitride NWHs is not solely determined by the band offsets and geometrical parameters such as ND height and NW diameter but also governed by a complex interplay between strain relaxation, polarization-induced internal electric fields and surface band bending. In order to elucidate the mechanisms of quantum confinement and in particular the effect of barrier composition on the strain distribution and transition energies, we have performed numerical simulations based on a 3-dimensional effective mass model for the self-consistent solution of the Schr\"odinger-Poisson equation in the NDs implemented in the \emph{nextnano}$^3$ software package \cite{nn3-website}. Due to the asymmetric nature of the investigated structures along the NW axis numerical simulations of the quantum confinement covered the whole 9-fold NWH in order to account for the complex strain distribution along the NW and to fully consider the impact of strain-induced effects. Recently, we have reported a comparison of single GaN/AlGaN ND, single NWH (with 9 NDs) and NWH ensemble PL measurements and have shown that the ND emission within one NW for $\text{[Al]}_\text{bar} = 0.14$ is dispersed by 45\,meV whereas the FWHMs of the emission of single NDs are between 2\,meV and 10\,meV \cite{Rigutti2010a}. 
The calculations presented here cover the dependence of the multiple ND emission properties on $\text{[Al]}_\text{bar}$ in the whole composition range $0 < x \leq 1$. The evolution of the emission energy and intensity with $\text{[Al]}_\text{bar}$ is analyzed and the resulting trends are compared to the experimental data. In particular, the origin of the non-monotonous relation of the ND emission energy on $\text{[Al]}_\text{bar}$ which has not been observed for other types of GaN/AlGaN heterostructures, is studied. 

The structural input parameters for the simulations were extracted from TEM-analysis. The NWs were simulated as hexagonal wires consisting of a GaN-base part (35\,nm in length and 30\,nm in diameter) followed by the GaN/AlGaN MQW system (thickness of barrier and NDs: 28\,ML and 7\,ML, respectively) and a 20\,nm cap region of the barrier material. The material parameters used for the simulations are summarized in Table~\ref{tab:Sim}. The presence of an AlGaN shell with identical composition as the barrier material was taken into account resembling a gradually increasing ND diameter from base (ND\,\#1) to top (ND\,\#9) as observed by TEM. For comparison, simulations assuming AlN as the shell material have been performed for intermediate $\text{[Al]}_\text{bar} = 0.3,\,0.4,\,\text{and}\,0.6$ where the strongest impact of an Al-rich shell is expected (not shown here, supplementary material). The lateral growth rate was parameterized as 11\% of the axial growth rate, linearly increasing with $\text{[Al]}_\text{bar}$. The volume of the GaN base region was kept constant for all simulated barrier compositions resulting in an increasing total NW diameter with higher $\text{[Al]}_\text{bar}$. A rectangular numerical mesh with a density of two mesh-points per nm in each direction was used.

% Table 2 ***************************************************************************************************

\subsubsection*{Strain Distribution}

For calculation of the strain distribution the integral elastic energy was minimized applying zero stress boundary conditions at the NW surface \cite{Povolotskyi2005} (achieved by defining a surrounding air cluster within \emph{nextnano}$^3$) and assuming fully coherent interfaces in accordance with theoretical considerations \cite{Glas2006}. 

Figure~\ref{fig:Fig07}a displays the $e_{zz}$ component ($z$ parallel to the polar c-axis)  of the strain tensor for the case of AlN barriers ($\text{[Al]}_\text{bar} = 1.0$) as a cross section along the NWH and shows a strong variation of the strain state inside the NDs as a consequence of the asymmetric structure. While ND\,\#1 (close to the base part) is strongly influenced by the GaN-base part and a thick AlN-shell resulting in moderate in-plane compressive strain, ND\,\#9 (at the top) is highly compressively strained  due to the AlN-cap layer and a small lateral shell thickness. This interplay results in strain variations along the ND-stack which sensitively depend on the barrier material and affirms the necessity to consider the total structure in the calculations.

% Figure 6 ***************************************************************************************************

Figure~\ref{fig:Fig07}b shows the variation of the $e_{zz}$-strain-component along the NW-axis for various $\text{[Al]}_\text{bar}$ which becomes more pronounced with increasing $\text{[Al]}_\text{bar}$. In particular, the c-lattice parameter in the GaN-base decreases considerably due to the vertical compression induced by the surrounding shell. Considering the gradient in shell thickness along the NW-base this can be regarded as the main reason for the broadening of the GaN-emission with increasing $\text{[Al]}_\text{bar}$, i.e.~with increasing shell thickness (cf. Figure~\ref{fig:Fig05}a).

\subsubsection*{Profile of the band edges}

For calculation of the resulting valence- and conduction band profiles, spontaneous polarization, strain-induced piezoelectric polarization and the effect of deformation potentials have been taken into account. 
The position of the Fermi level at the surface is determined by the distribution and density of surface states. However, the electronic structure of polar and non-polar AlGaN surfaces is not precisely known, particularly after the samples have been exposed to atmosphere and a native oxide layer as well as adsorbates are present on the surface \cite{Ambacher1996, Dong2006}. Here, we assume a pinning of the Fermi level at mid-gap of the lateral shell material, i.e.~the barrier material at all NW surfaces. This assumption is further justified by the weak dependence of the ND band profiles and the resulting transition energies on the position of the Fermi level pinning when, as in the present case, the conduction and valence bands of the GaN-NDs are sufficiently separated from the Fermi level \cite{Pfueller2010, Teubert2011} and the diameter of the NWs is small enough to consider them fully depleted \cite{Calarco2005}. We further assumed a residual doping concentration of $1 \times 10^{17}\,\text{cm}^{-3}$ and $1 \times 10^{16}\,\text{cm}^{-3}$ for the GaN and AlGaN-regions, respectively. However, as the NWs are fully depleted, the band profiles are hardly affected by surface band bending effects and also the precise doping concentration is not a key parameter. 

Figure~\ref{fig:Fig08} depicts lateral conduction band profiles obtained at the top of the respective NDs, evidencing the strong influence of the gradient of the lateral AlGaN shell thickness along the NW. With increasing shell thickness the shape of the confinement potential changes, as shown in Figure~\ref{fig:Fig08}a for ND\,\#2 and increasing $\text{[Al]}_\text{bar}$. While for a thin lateral shell ($\text{[Al]}_\text{bar} \leq 0.3$) the lateral confinement potential for electrons shows a minimum at the NW center in accordance with ref.~\cite{Rivera2007}, the curvature exhibits a non-monotonous behavior with ring-shaped minima close to the ND boundary for a thicker shell ($\text{[Al]}_\text{bar} \geq 0.4$). For higher $\text{[Al]}_\text{bar}$ this transition can be even found within the same NWH (cf.~Figure~\ref{fig:Fig08}b for $\text{[Al]}_\text{bar} = 0.6$). In that case NDs close to the GaN base (thick shell) are characterized by ring shaped minima while NDs close to the top typically present the U-shaped lateral confinement potential. For higher $\text{[Al]}_\text{bar}$ this transition is shifted towards higher ND numbers. 
The radial valence band profiles, also shown in Figure~\ref{fig:Fig08}a, show a corresponding behavior with a well pronounced maximum of the VB energy at the center of the ND for $\text{[Al]}_\text{bar} > 0.4$ which leads to a localization of holes in the center of the ND.
Intermediate concentrations between $\text{[Al]}_\text{bar} = 0.3$ and 0.4 represent a unique situation with a high number of NDs exhibiting flat band conditions (cf.~Figure~\ref{fig:Fig08}c), which would not be the case without the presence of a lateral shell.

% Figure 7 ****************************************************************************************************

This behavior is also reflected in the resulting one-particle eigenstates. In a situation of thin lateral shell electrons are located at the NW center and holes at the ND boundary whereas for a thicker shell the behavior is inverted. The flatband situation for concentrations of $\text{[Al]}_\text{bar} = 0.3 \ldots 0.4$ results in the largest overlap of the one-particle eigenstates of electrons and holes. 

\subsubsection*{Strain induced lateral electric fields in the NDs}

As for most of the investigated NWs the ND height of $d_\text{ND} = 1.7\,\text{nm}$ is smaller than the exciton Bohr-radius, the question whether to describe the ND luminescence as recombination of excitons or of separated electron-hole pairs is mainly determined by the strength of lateral electric fields. According to refs.~\cite{Dow1970, Blossey1970, Shokhovets2003} the critical field strength for exciton splitting in GaN is approximately 80\,kV/cm. Evaluation of the lateral band profiles (cf.~Figure \ref{fig:Fig08}) for all NDs and Al-contents by numerical differentiation reveals a general trend towards increasing lateral electric fields with increasing $\text{[Al]}_\text{bar}$, exceeding a value of 80\,kV/cm at an Al-concentration of approximately 70\% (upper limit). Above this concentration electron- and hole wave functions will be spatially separated, resulting in a decreasing oscillator strength and a suppression of the excitonic character of the ND luminescence. 

\begin{figure}[tb]
	\includegraphics[trim=0.0cm 0.0cm 0.0cm 0.0cm, width=8cm]{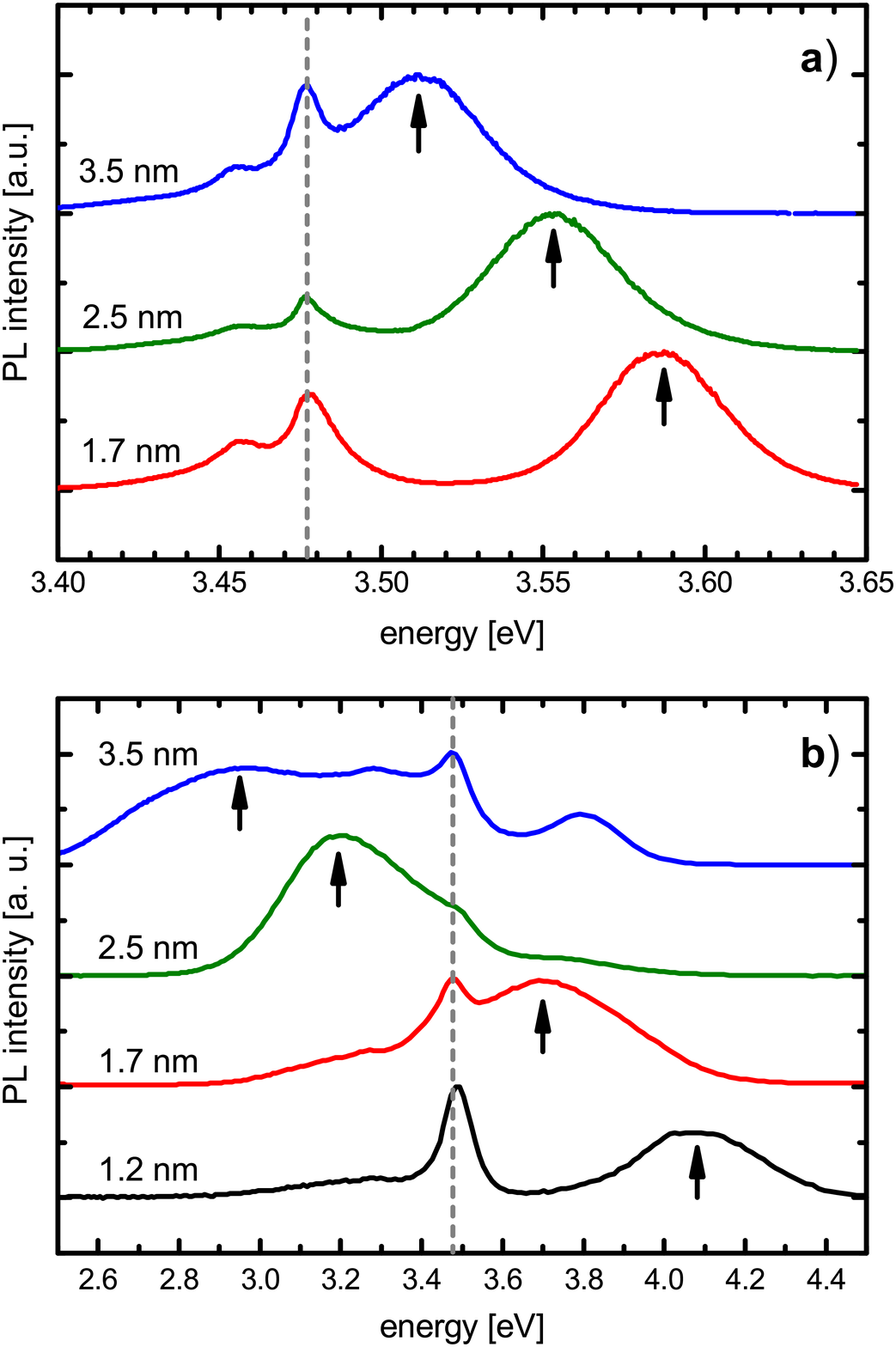}
    \caption{
(Color online)
Photoluminescence spectra of NWHs of different ND height $d_\text{ND}$ recorded at $T = 4\,\text{K}$. Individual spectra are vertically shifted for clarity. The ND related emission is indicated by arrows. The dotted line at 3.477\,eV indicates the NBE emission from the GaN base region.
a)	Three GaN/AlGaN NWH samples with $\text{[Al]}_\text{bar} = 0.14$ and $d_\text{ND} = 1.7\,\text{nm}$, 2.5\,nm, 3.5\,nm. 
b)	Four GaN/AlN NWH samples with $d_\text{ND} = 1.2\,\text{nm}$, 1.7\,nm, 2.5\,nm, 3.5\,nm.  
}
\label{fig:Fig04}
\end{figure}

\begin{figure}[tb]
	\includegraphics[trim=0.0cm 0.0cm 0.0cm 0.0cm, width=8cm]{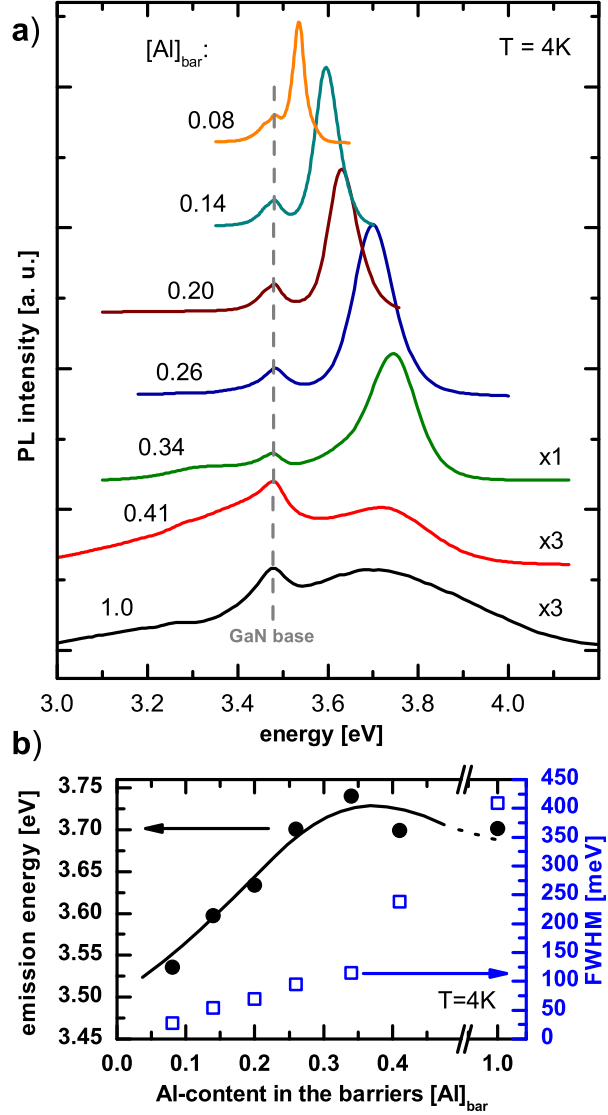}
    \caption{
(Color online) 
a) Low temperature photoluminescence spectra ($T = 4\,\text{K}$) of samples with a ND height of $d_\text{ND}=1.7\,\text{nm}$ and 7\,nm AlGaN barriers with Al content varying from $\text{[Al]}_\text{bar} = 0.08$ to $\text{[Al]}_\text{bar} = 1.0$. The vertical line at 3.48\,eV marks the position of the excitonic recombination in pure GaN NWs originating from the NW base part. Individual spectra are vertically shifted for clarity. 
b) Corresponding PL peak position and FHWM of the ND emission as a function of the Al content in the barriers $\text{[Al]}_\text{bar}$.  
}
\label{fig:Fig05}
\end{figure}

\begin{figure}[tb]
	\includegraphics[trim=0.0cm 0.0cm 0.0cm 0.0cm, width=8cm]{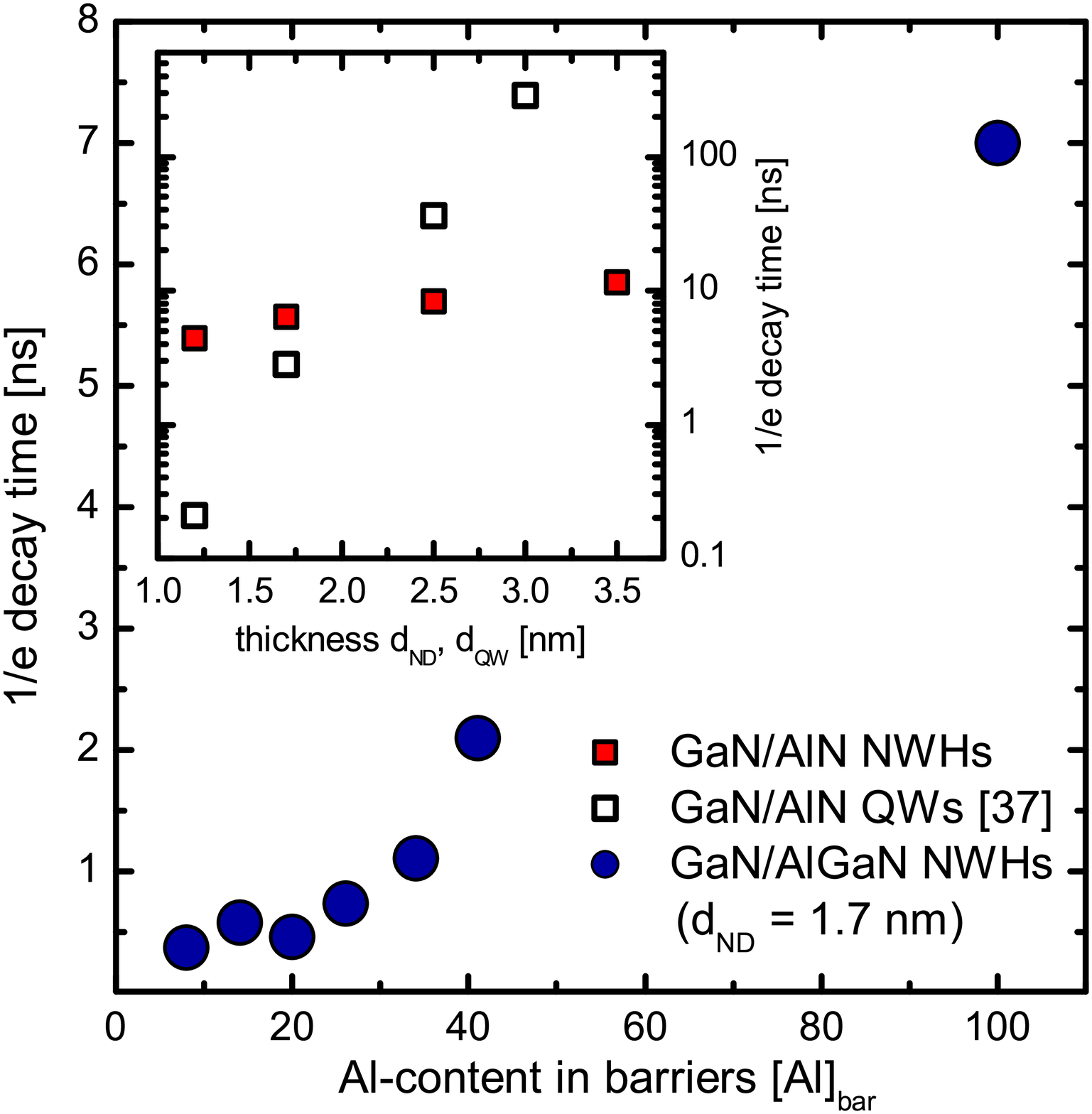}
    \caption{
(Color online) 
1/e decay times of the ND emission at $T = 10\,\text{K}$ for GaN/AlGaN NWH ensembles with $d_\text{ND}=1.7\,\text{nm}$ and different $\text{[Al]}_\text{bar}$ (blue filled circles) and (insert) for $\text{[Al]}_\text{bar} = 1.0$ as a function of $d_\text{ND}$ (red filled squares) in comparison with results from ref.~\cite{Renard2009a} for GaN/AlN QW structures with QW-width $d_\text{QW}$ (black open squares). 
}
\label{fig:Fig06}
\end{figure}

\begin{figure}[tb]
	\includegraphics[trim=0.0cm 0.0cm 0.0cm -0.5cm, width=8cm]{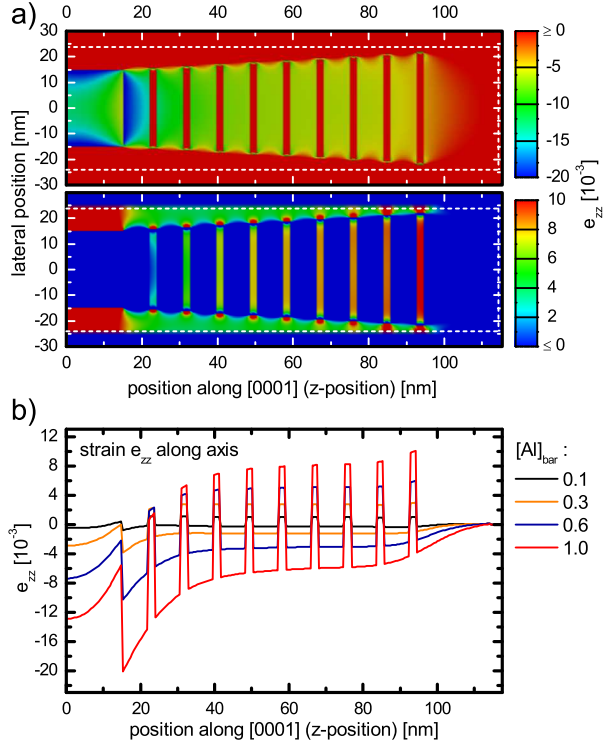}
    \caption{
(Color online)
a) Cross sectional view of the $e_{zz}$ component of the strain tensor for $\text{[Al]}_\text{bar} = 1.0$ (z-axis $\mid \mid $ [0001] growth direction). Both images show the same dataset. For clarity and contrast enhancement, the color scale in each image was restricted to show negative values of $e_{zz}$ (upper image) and positive values of $e_{zz}$ (lower image), respectively. Dotted white lines indicate the outer boundaries of the NW.
b)	Distribution of the $e_{zz}$ component of the strain tensor along the NW axis (along [0001]) for various $\text{[Al]}_\text{bar}$.
}
\label{fig:Fig07}
\end{figure}

\begin{figure}[tb]
	\includegraphics[trim=0.0cm 0.0cm 0.0cm -0.5cm, width=8cm]{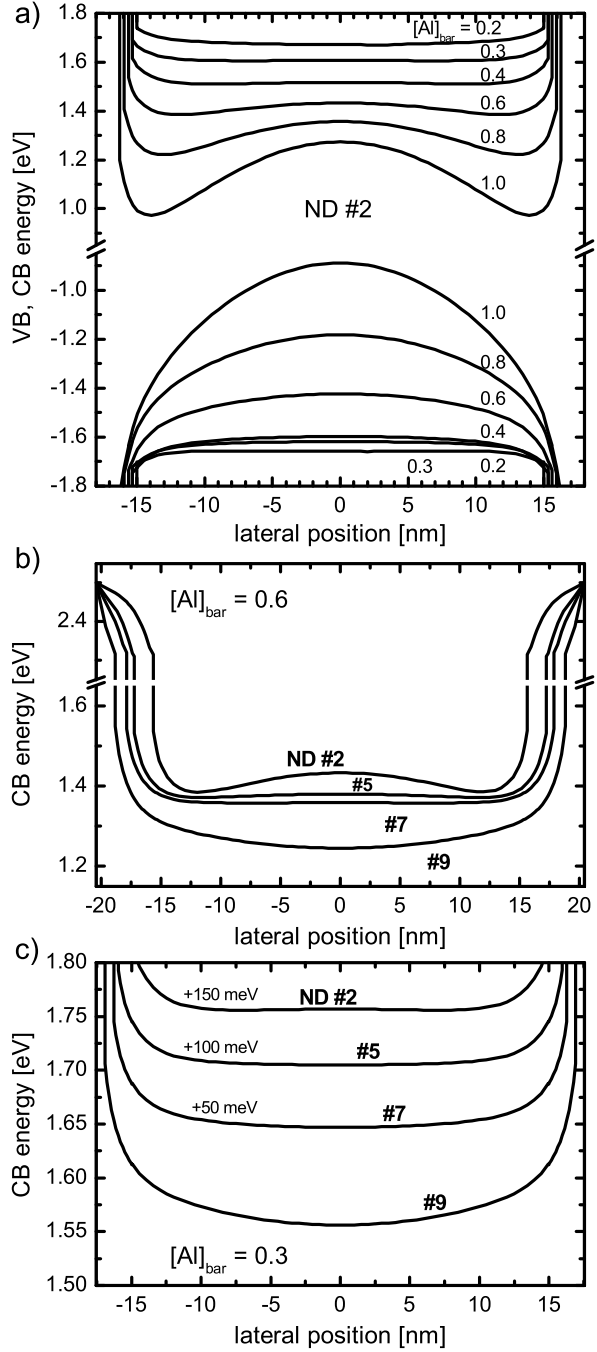}
    \caption{
(Color online)
Simulated lateral conduction band (valence band) profiles obtained at the top (bottom) of the respective NDs to represent the confinement potential for electrons (holes): a) for conduction band (CB) and valence band (VB) of ND\,\#2 and various Al-content in the barriers $\text{[Al]}_\text{bar}$; b) CB profiles for $\text{[Al]}_\text{bar} = 0.6$ and NDs\,\#2, \#5, \#7, and \#9; c) as b) for $\text{[Al]}_\text{bar} = 0.3$ (CB profiles are shifted for clarity as indicated). The position of the Fermi level defines 'zero' potential. 
}
\label{fig:Fig08}
\end{figure}

Furthermore, due to the lateral separation of electron and hole wave function, the effect of the axial internal electric field on the luminescence intensity via a separation along the polar growth direction according to the QCSE is weakened. Experimental evidence may be seen in the dependence of the luminescence intensity on $d_\text{ND}$ for a constant $\text{[Al]}_\text{bar}$ shown in Figure \ref{fig:Fig04}. The decrease of the relative emission intensity of the NDs with increasing ND height is more pronounced for $\text{[Al]}_\text{bar} = 0.14$ (Figure \ref{fig:Fig04}a) than for ND samples with AlN barriers (Figure \ref{fig:Fig04}b), despite the higher polarization-induced axial electric field in the latter and although the ND height exceeds the exciton Bohr radius. However, as PL-intensities have to be analyzed with great care, experimental evidence for the presence of strain induced lateral electric fields can only be obtained from time resolved PL spectroscopy. As shown in Figure \ref{fig:Fig06}, samples with $d_\text{ND} = 1.7\,\text{nm}$ show a strong increase of the decay times with increasing $\text{[Al]}_\text{bar}$ from values of a few hundred picoseconds at low $\text{[Al]}_\text{bar}$ up to 7\,ns for $\text{[Al]}_\text{bar} = 1.0$. We attribute this increase to the presence of strain induced lateral electric fields. This becomes more evident when considering the decay time measured for $\text{[Al]}_\text{bar} = 1.0$ and $d_\text{ND} = 1.2\,\text{nm}$ (insert of Figure\,\ref{fig:Fig06}). In that case the decay time of the ND emission (4\,ns) exceeds that of comparable QWs (200\,ps) by one order of magnitude. As GaN/AlGaN QW structures also show internal electric fields in polar direction (equivalent to the axial direction in NWs) but lack the presence of lateral electric fields due to symmetry, this observation can only be explained by exciton splitting due to the lateral electric field in the NDs. 

For $d_\text{ND} \geq 2.5\,\text{nm}$ the situation is different. Here the presence of dislocations and the accompanying increased probability for non-radiative recombination as well the reduction of the axial internal electric field, discussed below, is responsible for the smaller decay times compared to QWs.

\subsubsection*{Transition energies}

\begin{figure}[tb]
	\includegraphics[trim=0.0cm 0.0cm 0.0cm -0.5cm, width=8cm]{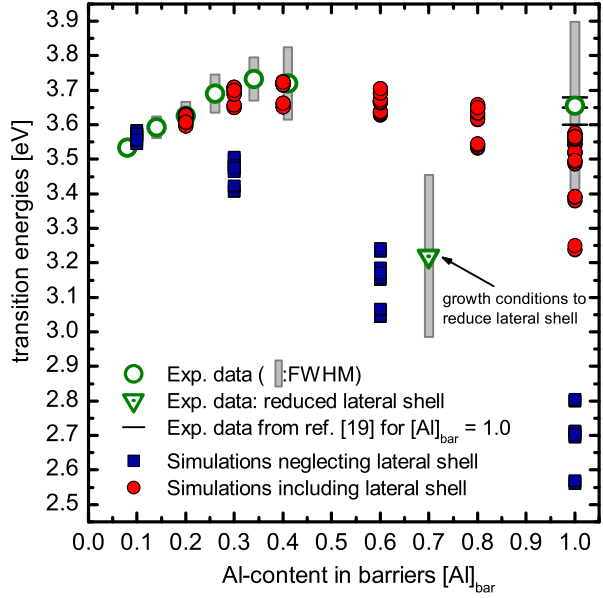}
    \caption{
(Color online)
Comparison of the calculated transition energies as a function of $\text{[Al]}_\text{bar}$ to the experimental data. PL-peak energies for samples containing a lateral shell (extracted from the spectra shown in Figure~\ref{fig:Fig05}) are given by green open circles; the sample grown under modified growth conditions to reduce the lateral growth is represented by a green open triangle. Grey bars indicate FWHM-values. Data from ref.~\cite{Zagonel2011} for $\text{[Al]}_\text{bar} = 1.0$ indicated by horizontal bars is presented for comparison.
Simulations taking the presence of a lateral shell into account are represented by red filled circles; blue filled squares indicate those neglecting lateral growth. Shown are transitions between electron and hole ground states for NDs\,\#2, \#5, \#7, and \#9. In case of a lateral shell and AlN barriers we considered all NDs\,\#2 - \#9. The individual energy values for a given Al-content correspond to the different NDs and give rise to an energetic dispersion. Without the presence of a shell (blue filled squares) we find the trend of decreasing transition energy from ND~\#2 (bottom) to ND~\#9 (top). However, in the case of an AlGaN shell a clear trend can only be given for $\text{[Al]}_\text{bar} > 0.3$, i.e.~the transition energies increase for the first NDs and saturate towards the top of the NW. For each ND the transition energy was corrected for excitonic effects (assuming an exciton binding energy of 40\,meV and an exciton localization energy of 10\,meV) in case the lateral electric fields did not exceed the value of 80\,kV/cm.
}
\label{fig:Fig09}
\end{figure}

To calculate the PL transition energies, the confined one-particle electron and hole states were calculated for ND\,\#2,~\#5,~\#7 and \#9 (NDs\,\#2 - \#9 for $\text{[Al]}_\text{bar} = 1.0$). Applying Dirichlet boundary conditions the Schr\"odinger equation was solved numerically for a 'quantum region' (defined within \emph{nextnano}$^3$) ranging from 3\,nm below to 4\,nm above the respective ND based on the full strain calculation described above. The transition energies obtained from the one particle eigenfunctions have been corrected for the exciton binding energy and exciton localization energy of 40\,meV and 10\,meV \cite{Grandjean1999a, Rigutti2010a}, respectively, in case that the maximum lateral electric field inside the respective ND does not exceed the critical value for exciton ionization of 80\,kV/cm. The obtained results, depicted in Figure~\ref{fig:Fig09} in comparison with the experimental PL results (green open circles, gray bars indicate FWHM), demonstrate that only by consideration of the lateral AlGaN shell (full red circles) agreement of numerical results and experimental data can be achieved. In particular, the position of the maximum emission energy between $\text{[Al]}_\text{bar} = 0.3 \ldots 0.4$ and the strength of the energetic dispersion with increasing $\text{[Al]}_\text{bar}$ are determined by this structural property. 

% Figure 8 *****************************************************************************************************

The strong impact of the lateral shell on the transition energies is demonstrated by comparing the results to those of simulations where the presence of the lateral shell was neglected (blue squares). In that case, the transition energy monotonously decreases with increasing $\text{[Al]}_\text{bar}$ in both qualitative and quantitative contradiction to the experimental data for structures characterized by a lateral shell as discussed here. The reliability of the numerical simulations is further corroborated by taking samples with reduced shell thickness (grown under modified growth conditions as described in section~\ref{ExpSample}) into account (cf.~Figure~\ref{fig:Fig01}d and green open triangle at $\text{[Al]}_\text{bar} = 0.72$ in Figure~\ref{fig:Fig09}). In that case, the experimental transition energy is significantly reduced and can be reasonably well described by the simulations neglecting the lateral shell.

% Figure 9 *********************************************************************************************************

For $\text{[Al]}_\text{bar} = 1.0$ the data for the transition energies of GaN/AlN ND samples from ref.~\cite{Zagonel2011} is shown, which also contained a lateral AlN shell. Although the exact shell thickness and the number of NDs in those samples were different the results of the calculations show good agreement to those data. These results further corroborate the importance of the lateral shell to influence the carrier confinement in the investigated NWHs. For $\text{[Al]}_\text{bar} = 1.0$ the experimental results for the transition energies are higher than the results of the calculations. This deviation can be attributed to the relaxation of NDs by formation of misfit dislocations and the resulting reduction of the piezoelectric polarization in the NDs that is not taken into account in the calculations. Calculations assuming a shell consisting of AlN showed worse agreement to the experimental data (cf.~supplementary material). In particular, a stronger energetic dispersion was found in that case which is not consistent with the experimental values for the FWHM. 
%This result backs up the TEM-results (cf.~section \ref{TEM}) regarding the shell material.

\subsubsection*{Polarization-induced axial electric fields in the NDs}

As already indicated in the discussion of Figure~\ref{fig:Fig04}, the presence of polarization-induced internal electric fields in quantum confined systems significantly influences the observable photoluminescence transition energies. Since two-dimensional quantum well structures exhibit a laterally homogeneous strain distribution the strength of the internal electric field in such structures can be extracted from the dependence of the quantum well emission energy on the well thickness \cite{Grandjean1999a}.

In the previous paragraphs it was shown that stress relaxation at free lateral surfaces of GaN/AlGaN NWHs and the presence of a lateral shell results in a radial strain distribution that strongly modifies the lateral band profile. An increase in the ND height significantly modifies this strain distribution, the resulting lateral confinement potential as well as the related eigenvalues and the distribution of one-particle electron and hole wave functions.

As a consequence, the extraction of the electric field strength from the dependence of the emission energy on the ND height is not straight forward and solely possible from advanced modeling, in particular in the case of high $\text{[Al]}_\text{bar}$, where the increased lateral internal field strength causes separation of electrons and holes. 

\begin{figure}[tb]
	\includegraphics[trim=0.0cm 0.0cm 0.0cm -0.5cm, width=8cm]{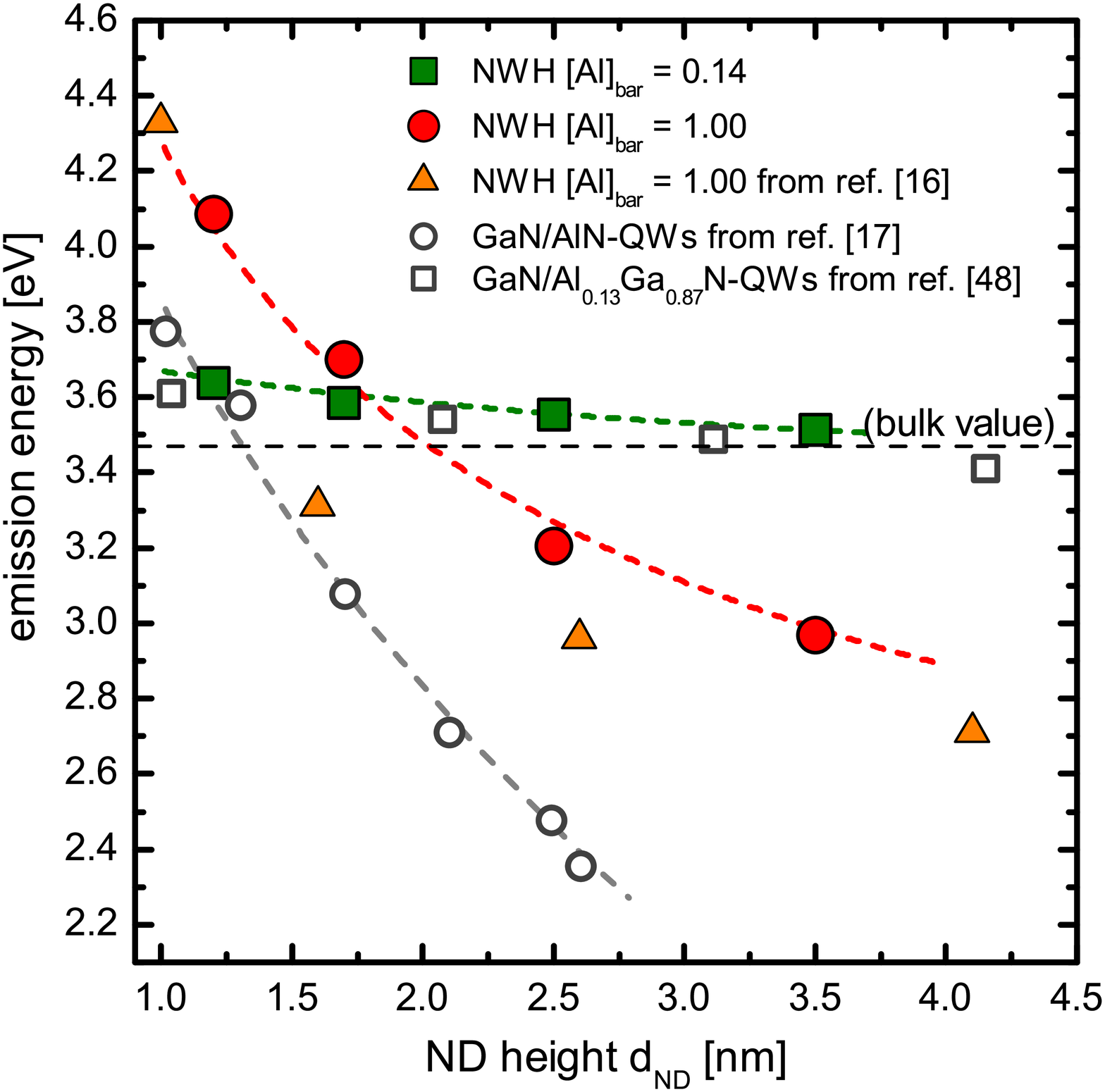}
    \caption{
(Color online)
Evolution of the emission energy with well thickness for samples with $\text{[Al]}_\text{bar} = 1.0$ (red filled circles) and $\text{[Al]}_\text{bar} = 0.14$ (green filled squares). For comparison, data from ref.~\cite{Renard2009} obtained from GaN/AlN-NWHs at room temperature (orange filled triangles), as well as low-temperature data from GaN quantum wells in Al$_{0.13}$Ga$_{0.87}$N barriers \cite{Grandjean1999} (open squares) and AlN barriers \cite{Adelmann2003} (open circles) is included. Dashed lines serve as guide to the eye.
}
\label{fig:Fig10}
\end{figure}

In Figure~\ref{fig:Fig10} the measured PL peak positions of ND samples with $\text{[Al]}_\text{bar} = 0.14$ (cf.~Figure~\ref{fig:Fig04}a) as well as respective results for AlN barriers (cf.~Figure~\ref{fig:Fig04}b) are displayed as a function of the ND height. For comparison, also results from Renard et al.~\cite{Renard2009} on GaN/AlN single ND-NWHs with similar barrier width and results obtained for quantum well structures for both Al$_{0.13}$Ga$_{0.87}$N \cite{Grandjean1999} and AlN-barriers \cite{Adelmann2003} are shown. The results from Renard et al.~show a similar trend as observed for the structures here. The consistently lower emission energies are due to the fact that in the structures investigated in that work only single NDs were used, which in our simulations give rise to the lowest emission energy (ND~\#1). 

Good agreement is also found for low $\text{[Al]}_\text{bar}$ with the results from GaN/Al$_{0.13}$Ga$_{0.87}$N QWs (data from ref.~\cite{Grandjean1999}). However, in the case of AlN barriers the dependence of the emission energy on well thickness is significantly weaker in NDs than in two dimensional GaN/AlGaN quantum wells. Also for GaN quantum dot structures in AlN matrix a decrease of the emission energy to 1.8\,eV for a dot height of 3.5\,nm was observed \cite{Bretagnon2006}.

The apparent reduction of the internal electric field in NDs compared to quantum wells was also reported by Renard et al. \cite{Renard2009} for GaN/AlN NWHs and by Zamifirescu et al. \cite{Zamfirescu2005a} where an internal electric field of 0.4\,MV/cm, one third of the theoretically expected value, was estimated for NDs with Al$_{0.28}$Ga$_{0.72}$N barriers. Mojica et al. explained the reduction of the QCSE by screening effects \cite{Mojica2010} originating from the presence of a two dimensional electron gas below the AlGaN barrier and a positive surface charge at the top NW-surface. However, the influence of this effect decreases with increasing distance $d$ between the two dimensional electron gas and the polar surface \cite{Mojica2010} and can be regarded as not significant in the present 9-fold NWHs (with $d = 100\,\text{nm}$). As no qualitative difference between our data ($d = 100\,\text{nm}$) and the experimental results from ref.~\cite{Renard2009} ($d = 10\,\text{nm}$) are found, the importance of the screening effect put forward in ref.~\cite{Mojica2010} cannot be confirmed. According to the slope of conduction and valence band in axial direction obtained in the present simulations (for NDs with $d_\text{ND} = 1.7\,\text{nm}$ shown above) the internal electric field along the polar direction in the center (border) of the NDs can be extracted to vary between 900\,kV/cm (700\,kV/cm) for Al$_{0.05}$Ga$_{0.95}$N barriers and 10\,MV/cm (8\,MV/cm) for AlN barriers in pseudomorphic structures. 

This apparent contradiction between the present results, the values expected from other experimental reports in literature and the simulations indicates that the c-lattice parameter inside the GaN NDs, that determines the magnitude of the piezoelectric polarization, is smaller than it would be expected for full pseudomorphic growth between the AlGaN barriers. The formation of single misfit dislocations in the NDs, described for AlN barriers in Figure~\ref{fig:Fig03} and the related discussion, can explain this behavior. The presence of two dislocations in the cross section of a NW with a diameter of 40\,nm corresponds to a dislocation density of $5 \times 10^{10}\,\text{cm}^{-2}$, i.e.~they can cause almost complete relaxation of strain and a decrease of the c-lattice parameter compared to pseudomorphic growth of GaN on AlGaN implying a strong reduction of the piezoelectric polarization. The simultaneous growth of the lateral AlGaN shell can further promote this effect by exerting compressive strain to the outer edge of the NDs, where the electron wave function is preferentially localized for high $\text{[Al]}_\text{bar}$. As a consequence, the polarization-induced internal electric fields in the NDs observable by evaluation of the PL data for different ND heights is significantly reduced. This may also explain the underestimation of the emission energy for AlN barriers by the numeric simulations shown in Figure~\ref{fig:Fig09}.

Recently, Landr\'e at al.~have reported purely elastic strain relaxation without formation of dislocations of 2\,nm GaN NDs between 2.3\,nm AlN barriers \cite{Landre2010}. On the other hand, in ref.~\cite{Renard2009} a significant decrease of the internal electric field in AlN/GaN NDs with 10\,nm AlN barriers was observed for ND height above 2\,nm by PL measurements. In contrast to the first case the results of the latter report could be well explained by strain relaxation due to misfit dislocations, as for 10\,nm AlN barriers, comparable to the structures investigated in this work, the compressive strain in the NDs would be increased almost by a factor of 5 compared to the structures with 2.3\,nm barriers analyzed in ref.~\cite{Landre2010}, if an effective medium model for the strain is assumed. This is further confirmed by the comparison of the respective data to the results of this study in Figure~\ref{fig:Fig10}. 

In addition full electromechanical coupling, recently proposed in different works \cite{Jogai2003, Christmas2005, Willatzen2006} has to be considered to precisely describe the polarization-induced internal electric fields in NWHs. This implies that, due to the piezoelectric properties, the polarization-induced internal electric fields cause a compression along the polar direction that counteracts the extension due to compressive in-plane strain. In In$_{0.2}$Ga$_{0.8}$N/GaN quantum well structures this effect has been found to cause a reduction of the internal fields by 2\% \cite{Christmas2005}. In refs.~\cite{Jogai2003, Willatzen2006} electromechanical coupling due to piezoelectric and spontaneous polarization on the strain in GaN/AlGaN heterostructures has been found to reduce the strain caused by the pseudomorphic growth significantly ($\sim 10\%$). This effect can be screened by the presence of free carriers or accumulated carriers in a two dimensional electron gas. However, in the present case, the NDs are embedded in NWs with a diameter of around 50\,nm, which according to ref.~\cite{Calarco2005}, can be considered as fully depleted due to surface band bending \cite{Pfueller2010, Teubert2011}. Thus, a reduction of the above mentioned electromechanical coupling effects due to screening of polarization-induced internal electric fields can be neglected.

\section{Conclusions}

In conclusion, axial GaN/AlGaN nanowire heterostructures including nanodiscs (NDs) were grown in the whole composition range ($\text{[Al]}_\text{bar} = 0.08 \ldots 1.0$) and investigated with respect to their structural and optical properties, in particular the quantum confinement in the NDs. The evolution of the ND emission energy with ND height $d_\text{ND}$, studied by low temperature photoluminescence (PL) revealed a good agreement with comparable two dimensional quantum well (QW) structures for low $\text{[Al]}_\text{bar}$, while the redshift observed in GaN/AlN-NWHs due to the quantum confined Stark effect is much less pronounced than for GaN/AlN-QWs. This effect is mainly attributed to strain relaxation as a result of the presence of misfit dislocations at the ND interface for $d_\text{ND}\geq 3\,\text{nm}$. Furthermore, the dependence of the PL properties on $\text{[Al]}_\text{bar}$, which showed a non-monotonous dependence with a maximum at $\text{[Al]}_\text{bar} = 0.3$ was analyzed by comparison of experiment and three dimensional numerical simulations with structural properties obtained by TEM as input parameters. The calculations indicate a strong influence of the lateral shell thickness on the confinement potential. Intermediate $\text{[Al]}_\text{bar} = 0.3 \ldots 0.4$ result in the highest relative emission intensities due to the lowest lateral electric fields corresponding to a flatband situation inside the NDs. For higher $\text{[Al]}_\text{bar}$ the lateral electric fields lead to a suppression of the excitonic character of the emission confirmed by TR-PL experiments. The dependence of the transition energy on $\text{[Al]}_\text{bar}$ observed in the PL experiments can only be explained when the presence of the lateral shell is taken into account. The simulations also show an increasing in-wire energetic dispersion with $\text{[Al]}_\text{bar}$ which is interpreted as an important contribution to the observed increase in FWHM of the ND emission.

\section*{Acknowledgements}
JT, while listed as second author, shares first authorship on this article. The authors gratefully acknowledge financial support via Deutsche Forschungsgemeinschaft (Grant No.: Ei 518/2 (NAWACS)) and by the EU within the Project DOTSENSE (STREP 224212). Support by the Spanish Government projects Consolider Ingenio 2010 CSD2009 00013 IMAGINE and CSD2009 00050 MULTICAT as also acknowledged. JA acknowledges funding from the Spanish CSIC project NEAMAN and the MICINN project MAT2010-15138 (COPEON). The authors would like to thank the TEM facilities in Serveis Cientificotècnics from Universitat de Barcelona. JT acknowledges support from the JLU via a Just'us fellowship.

%\bibliographystyle{unsrt}

%\bibliographystyle{phaip}

%\bibliography{L:/Literatur_AG-Eickhoff}
%\bibliography{D:/Papers/Literatur_Server/Literatur_AG-Eickhoff}

% from bbl file*******************************************************************************************************

%*********************************************************************************************************************

\newpage

\begin{table*}
\begin{tabular*}{14cm}[t]{|c@{\extracolsep{\fill}}|c|c|c|c|c|c|}
\hline
\parbox{1.5cm}{$T_\text{Al} [^\circ \text{C}]$} {\rule[0mm]{0mm}{6mm}} &
\parbox{1.5cm}{$T_\text{Ga} [^\circ \text{C}]$} & 
\parbox{2.0cm}{BEP$_\text{Al}$ [mbar]} & 
\parbox{2.0cm}{BEP$_\text{Ga}$ [mbar]} & 
\parbox{1.5cm}{[Al]$_\text{bar}$}  & 
\parbox{2.0cm}{AlGaN peak energy [eV] *)} & 
\parbox{2.0cm}{[Al]$_\text{bar}$ from PL peak pos.} \\
\hline
\hline
1020 {\rule[0mm]{0mm}{4mm}} &      & $1.40 \times 10^{-8}$ &                       & 0.03 & 3.52 & 0.03 \\ \cline{1-1} \cline{3-3} \cline{5-7}
1056 {\rule[0mm]{0mm}{4mm}} &      & $3.30 \times 10^{-8}$ &                       & 0.08 & 3.60 & 0.08 \\ \cline{1-1} \cline{3-3} \cline{5-7}
1074 {\rule[0mm]{0mm}{4mm}} &      & $4.70 \times 10^{-8}$ &                       & 0.11 & -    & -    \\ \cline{1-1} \cline{3-3} \cline{5-7}
1092 {\rule[0mm]{0mm}{4mm}} &      & $6.50 \times 10^{-8}$ &                       & 0.14 & 3.67 & 0.13 \\ \cline{1-1} \cline{3-3} \cline{5-7}
1110 {\rule[0mm]{0mm}{4mm}} & 1012 & $9.70 \times 10^{-8}$ & $3.90 \times 10^{-7}$ & 0.20 & -    & -    \\ \cline{1-1} \cline{3-3} \cline{5-7}
1130 {\rule[0mm]{0mm}{4mm}} &      & $1.40 \times 10^{-7}$ &                       & 0.26 & 3.98 & 0.28 \\ \cline{1-1} \cline{3-3} \cline{5-7}
1150 {\rule[0mm]{0mm}{4mm}} &      & $2.00 \times 10^{-7}$ &                       & 0.34 & -    & -    \\ \cline{1-1} \cline{3-3} \cline{5-7}
1170 {\rule[0mm]{0mm}{4mm}} &      & $2.70 \times 10^{-7}$ &                       & 0.41 & -    & -    \\ \cline{1-1} \cline{3-3} \cline{5-7}
1185 {\rule[0mm]{0mm}{4mm}} &      & $3.50 \times 10^{-7}$ &                       & 1.00 & -    & -    \\ \hline
1160 {\rule[0mm]{0mm}{4mm}} & 965  & $2.30 \times 10^{-7}$ & $0.88 \times 10^{-7}$ & 0.72 & -    & -    \\ \hline
\multicolumn{7}{l}{
\parbox[l][][l]{12.0cm}{*) PL-energy related to AlGaN barrier material obtained for reference samples with long cap of barrier material.}} \\
\end{tabular*}
 \caption{Al ($T_\text{Al}$) and Ga ($T_\text{Ga}$) effusion cell temperatures, corresponding beam equivalent pressures (BEP$_\text{Al}$ and BEP$_\text{Ga}$), and barrier composition [Al]$_\text{bar}$ of the investigated samples. The Al content was estimated by two methods: (i) by the flux ratio of BEP$_\text{Ga}$ and BEP$_\text{Al}$ without taking Ga desorption into account and (ii) from PL measurements on reference samples (GaN NWs with approx. 100\,nm AlGaN on top; no NDs). For the determination of [Al]$_\text{bar}$ from the PL peak energy, a bowing parameter of $b = 1.3\,\text{eV}$ was chosen according to ref.~\cite{Angerer1997}.}
 \label{tab:GrowthDetails}
\end{table*}

\begin{table*}
\begin{tabular*}{12cm}[t]{|l@{\extracolsep{\fill}}|l|}
\hline
\parbox{2.0cm}{Parameter} {\rule[0mm]{0mm}{6mm}} &
\parbox{6.0 cm}{Value as a function of  $x=\text{[Al]}_\text{bar}$} \\
\hline
\hline
\hspace{0.5 cm} $a$ [nm]                           {\rule[0mm]{0mm}{4mm}} &    $0.3112 x + 0.3189 (1-x)$   \\ \hline
\hspace{0.5 cm} $c$ [nm]                           {\rule[0mm]{0mm}{4mm}} &    $0.4982 x + 0.5185 (1-x)$   \\ \hline
\hspace{0.5 cm} $m^*_\text{perpendicular}$         {\rule[0mm]{0mm}{4mm}} &    $0.32 x + 0.206 (1-x)$   \\ \hline
\hspace{0.5 cm} $m^*_\text{parallel}$              {\rule[0mm]{0mm}{4mm}} &    $0.30 x + 0.202 (1-x)$   \\ \hline
\hspace{0.5 cm} $E_\text{C}\,^{*)}$                {\rule[0mm]{0mm}{4mm}} &    $2.7997 x + 4.7245 (1-x) - 0.7 x(1-x)$   \\ \hline
\hspace{0.5 cm} $E_\text{V,A}\,^{*)}$              {\rule[0mm]{0mm}{4mm}} &    $-0.7327 x - 1.4133 (1-x)$   \\ \hline
\hspace{0.5 cm} $C_{11}$                           {\rule[0mm]{0mm}{4mm}} &    $396 x + 390 (1-x)$    \\ \hline
\hspace{0.5 cm} $C_{12}$                           {\rule[0mm]{0mm}{4mm}} &    $137 x + 145 (1-x)$   \\ \hline
\hspace{0.5 cm} $C_{13}$                           {\rule[0mm]{0mm}{4mm}} &    $108 x + 106 (1-x)$   \\ \hline
\hspace{0.5 cm} $C_{33}$                           {\rule[0mm]{0mm}{4mm}} &    $373 x + 398 (1-x)$   \\ \hline
\hspace{0.5 cm} $C_{44}$                           {\rule[0mm]{0mm}{4mm}} &    $116 x + 105 (1-x)$   \\ \hline
\hspace{0.5 cm} $e_{33}$\,[C/m$^2$]                {\rule[0mm]{0mm}{4mm}} &    $1.79 x + 1.27 (1-x)$   \\ \hline
\hspace{0.5 cm} $e_{31}$\,[C/m$^2$]                {\rule[0mm]{0mm}{4mm}} &    $-0.50 x - 0.35 (1-x)$   \\ \hline
\hspace{0.5 cm} $e_{15}$\,[C/m$^2$]                {\rule[0mm]{0mm}{4mm}} &    $-0.48 x - 0.30 (1-x)$   \\ \hline
\hspace{0.5 cm} $P_{\text{Sp}}$\,[C/m$^2$]         {\rule[0mm]{0mm}{4mm}} &    $-0.0340 x - 0.0900 (1-x) - 0.021 x(1-x)$ \\ \hline
\hspace{0.5 cm} $a_{1}$ [eV]                       {\rule[0mm]{0mm}{4mm}} &    $-3.4 - 4.9 (1-x)$   \\ \hline
\hspace{0.5 cm} $a_{2}$ [eV]                       {\rule[0mm]{0mm}{4mm}} &    $-11.8 - 11.3 (1-x)$   \\ \hline
\hspace{0.5 cm} $D_{1}$ [eV]                       {\rule[0mm]{0mm}{4mm}} &    $-17.1 - 3.7 (1-x)$   \\ \hline
\hspace{0.5 cm} $D_{2}$ [eV]                       {\rule[0mm]{0mm}{4mm}} &    $7.9 + 4.5 (1-x)$   \\ \hline
\hspace{0.5 cm} $D_{3}$ [eV]                       {\rule[0mm]{0mm}{4mm}} &    $8.8 + 8.2 (1-x)$   \\ \hline
\hspace{0.5 cm} $D_{4}$ [eV]                       {\rule[0mm]{0mm}{4mm}} &    $-3.9 - 4.1 (1-x)$   \\ \hline
\hspace{0.5 cm} $D_{5}$ [eV]                       {\rule[0mm]{0mm}{4mm}} &    $-3.4 - 4.0 (1-x)$   \\ \hline
\hspace{0.5 cm} $D_{6}$ [eV]                       {\rule[0mm]{0mm}{4mm}} &    $-3.4 - 5.1 (1-x)$   \\ \hline
\hspace{0.5 cm} donor energy $E_\text{d}$\,[meV]   {\rule[0mm]{0mm}{4mm}} &    $20$   \\ \hline

\multicolumn{2}{l}{
\parbox[l][][l]{12.0cm}{*) with respect to the standard hydrogen electrode}} \\
\end{tabular*}
 \caption{Material parameters for Al$_x$Ga$_{1-x}$N used in the \emph{nextnano}$^3$-simulations according to ref.~\cite{Vurgaftman2003}.}
 \label{tab:Sim}
\end{table*}

\begin{figure}[tb]
\begin{flushleft}
\bf{Supplementary material: \\ 	}
\end{flushleft}

\includegraphics[trim=0.0cm 0.0cm 0.0cm -0.5cm, width=8cm]{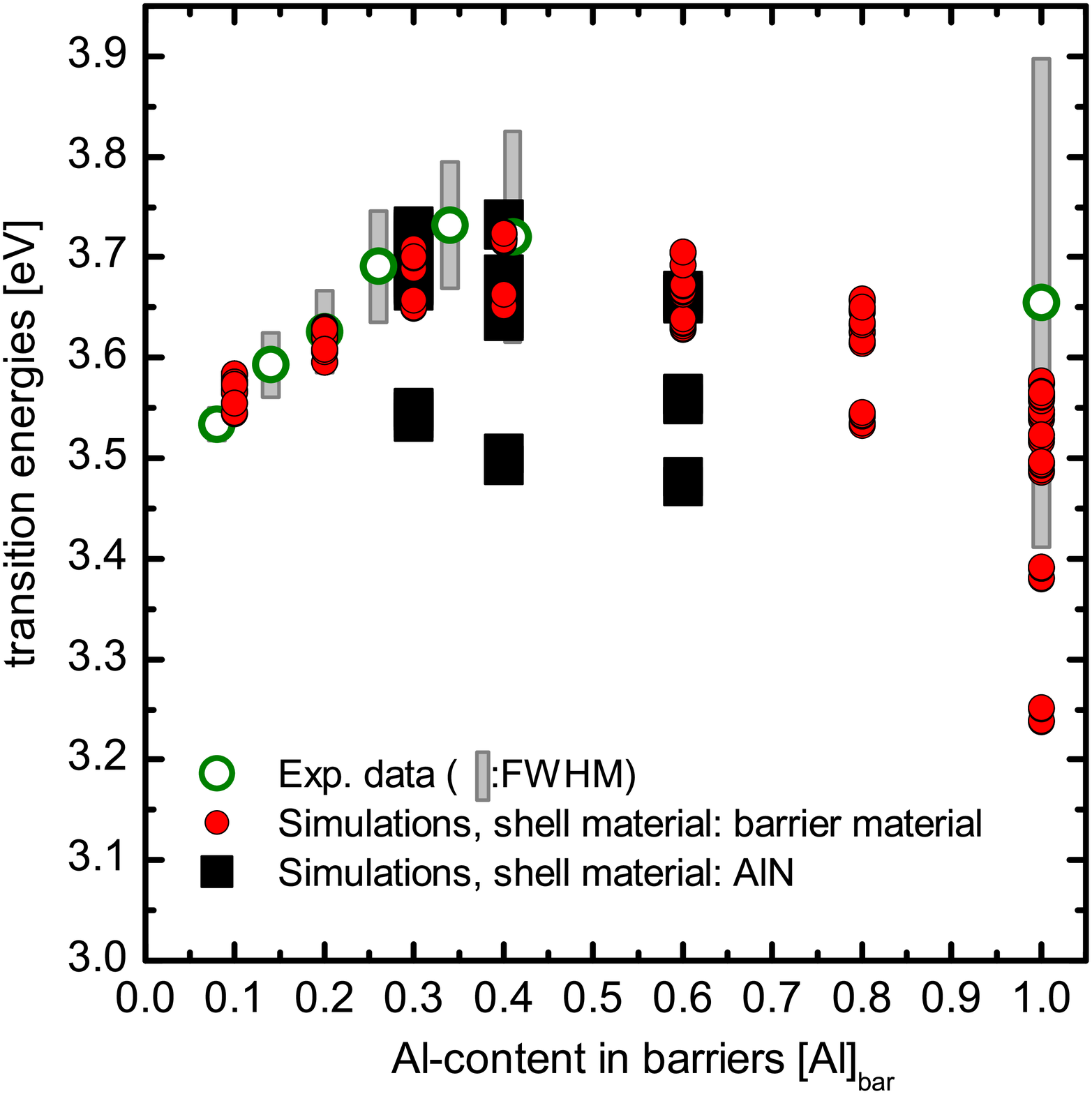} \\

Comparison of calculated transition energies as a function of $\text{[Al]}_\text{bar}$ to the experimental data. The data represented by red circles was obtained assuming that the composition of the shell is equal to that of the barrier material. Simulations assuming an AlN-shell are given by black squares. In the latter case much stronger dispersion is found which is not justified by the experimental FWHM.

%\label{fig:FigSup}
\end{figure}

%\begin{thebibliography}{99}

%\end{thebibliography}

\end{document}